\documentclass[11pt,a4paper]{article}
\usepackage[T1]{fontenc}

\usepackage[a4paper,top=2cm,bottom=2cm,left=2cm,right=2cm,marginparwidth=1.75cm]{geometry}

\usepackage{amsmath}
\usepackage{amssymb}
\usepackage{graphicx}
\usepackage{hyperref}
\usepackage{booktabs}
\usepackage{diagbox}
\usepackage{multirow}
\usepackage{enumitem}
\usepackage{xcolor}
\usepackage{amsthm}
\usepackage{caption}
\usepackage{subcaption}
\usepackage{setspace}
\usepackage{apacite}
\usepackage{parskip}
\usepackage{algorithm}
\usepackage{algpseudocode}
\usepackage{tabularray}

\usepackage{lipsum} 

\usepackage{indentfirst}

\usepackage{xcolor}

\usepackage[affil-it]{authblk}

\usepackage{natbib}

\newcommand{\gd}{\hat{\lambda}}
\newcommand{\lam}[1]{\lambda_{#1, r_l}}
\newcommand{\lambdas}{\lam{1}, \ldots, \lam{k}}
\newcommand{\prob}{\mathbb{P}}

\newcommand{\algopoints}[2]{ #1: \footnotesize \textit{#2} \normalsize}
\newcommand{\beti}{\beta_{t, i}}
\newcommand{\betrl}{\beta_t^{r_l}}
\newcommand{\lambdarl}{\lambda^{r_l}}

\providecommand{\keywords}[1]
{
  \small	
  \textbf{\textit{Keywords---}} #1
}

\title{Adaptive Optimisation of Ride-Pooling Personalised Fares in a Stochastic Framework}

\author[1,2]{Micha\l~Bujak*}
\author[1]{Rafa\l~Kucharski \footnote{Email addresses for authors: Micha\l~ Bujak (corresponding): michal.bujak@doctoral.uj.edu.pl; Rafa\l~ Kucharski: rafal.kucharski@uj.edu.pl.}}

\affil[1]{Faculty of Mathematics and Computer Science, Jagiellonian University, ul. Prof. S. \L ojasiewicza 6, 30-348 Krak\'ow}
\affil[2]{Doctoral School of Exact and Natural Sciences, Jagiellonian University, ul. Prof. S. \L ojasiewicza 11, 30-348 Krak\'ow}

\date{\vspace{-5ex}}

\begin{document}
\maketitle

\section*{Abstract}
Ride-pooling systems, to succeed, must provide an attractive service, namely compensate perceived costs with an appealing price. 
However, because of a strong heterogeneity in a value-of-time, each traveller has his own acceptable price, unknown to the operator. 
Here, we show that individual acceptance levels can be learned by the operator (over $90\%$ accuracy for pooled travellers in $10$ days) to optimise personalised fares. 
We propose an adaptive pricing policy, where every day the operator constructs an offer that progressively meets travellers' expectations and attracts a growing demand.
Our results suggest that operators, by learning behavioural traits of individual travellers, may improve performance not only for travellers (increased utility) but also for themselves (increased profit). 
Moreover, such knowledge allows the operator to remove inefficient pooled rides and focus on attractive and profitable combinations.

\keywords{ride-pooling, behavioural heterogeneity, pricing, Bayesian inference}

\section{Introduction}
Ride-pooling is an on-demand urban mobility service where travellers can be matched into shared rides. 
Compared to standard taxi (ride-hailing), travellers experience longer travel time (other passengers' pick-ups and drop-offs) and discomfort due to the sharing. 
Those negative effects are compensated with a lower fare expressed via a sharing discount.

The inconvenience experienced by a traveller relies heavily on his behavioural traits, mainly the value-of-time (\cite{zwick2022ride}, \cite{alonso2020value}). 
Some groups of people appreciate mostly shorter travel time, while others gladly sacrifice a quickness of a trip for monetary savings.
Therefore, the sharing offer to be appreciated must be precisely tailored towards individual expectations.
The attractiveness depends on how well the operator predicts the latent traits of individual travellers, mainly their value-of-time.
If they are well predicted, the operator may offer individual prices in an optimised way, which not only increases the travellers' satisfaction but also system performance and platform profit (as we show in \cite{bujak2024balancing}).
The operator tries to construct a profitable offer, yet he has only a few attempts before an unsatisfied client leaves the system. 
It is an asymmetrical relation where it is way easier to loose the client than gain his loyalty (\cite{chen2007cross}).

We speculate that neglecting the individualism is the main reason why the ride-pooling systems often fail before reaching critical mass.
Therefore, we develop an adaptive pricing policy. 
We propose a sequential learning method where, in a sequence of repeated days, the operator learns the individual value-of-time in a heterogeneous population.
Each day, he observes travellers' choices (who accept or reject shared rides) and, by applying Bayesian inference, estimates individual behavioural classes. 
With observed decisions, the operator's knowledge grows and he improves his understanding of one's expectations. 
On the daily basis, he fits personalised fares tailored with his current estimates of clients' value-of-time and constructs an offer which is attractive for travellers, profitable for him and attracts a growing demand.

We develop a stochastic framework, where travellers recognise offered shared rides as attractive or unattractive according to the proposed individualised fare and their actual behavioural class (latent to the operator).
To find the optimal fares, the operator devises all accept/reject decisions, along with their respective probability.
To combine traveller's perspective (acceptance probability) with service profitability, we introduce the expected profit measure.
For our policy to achieve long-term service goals, we track clients' satisfaction which reflects how likely a traveller is to join the service the next day.
Based on the satisfaction, we introduce an attraction value.
It accounts for the fact that satisfied travellers are more likely to rejoin the service while discouraged clients likely drop out of the system. 
Hence, to tailor the optimal personalised fares, we assess simultaneously current value (measured with the expected profit) and demand acquisition perspectives (approximated with the attraction value).
As we show in the ablation study, each component plays a vital role in the construction of an efficient ride-pooling pricing policy.

In our framework, we analyse a ride-pooling service during multiple consecutive days.
We propose a pricing policy which is adopted by the operator.
We assume that, for each potential traveller, the operator knows their initial likelihood of joining the service.
The population is divided into classes where each exhibits a different value-of-time distribution as highlighted in empirical studies (\cite{alonso2021determinants}, \cite{zwick2022ride}). 
On the first day of the service, the operator knows only the general distribution and not the individual preferences.
He conducts a discount optimisation process, where he considers both the current value and the future impact on the system (attraction value). 
Next, for an established offer, we sample travellers' decisions according to their true, latent to the operator, value-of-time.
The operator, based on the observed accept/rejection choices, applies the Bayesian inference to learn the behavioural classes of individual travellers. 
After several iterations (days), the estimates of the individual value-of-time are narrow. 
It allows to fit an offer which is both more appealing to travellers and more profitable for the operator. 

Our adaptive pricing policy uncovers the potential laying in the population heterogeneity.
Learning individual traits proves to be crucial in the construction of an optimal offer. 
Narrowing estimates of the value-of-time and including the attraction value allows the operator to attract more travellers and reach a critical mass required for an efficient pooling service. 
As we show in the numerical study, the expected profit of the operator grows in time and its variance decreases.
The operator learns the individual expectations, offers more shared rides, and achieves a higher acceptance rate. 
Also, travellers' satisfaction increases and they more likely join the service each day.
Benchmarking against simplified pricing policies highlights the importance of each of the introduced individual components of our model.

\subsection{Background}
The ride-pooling service, compared to a standard ride-hailing, offers mileage and fleet size reduction. \cite{engelhardt2019quantifying} and \cite{martinez2015agent} analyse benefits when ride-pooling is introduced on top of an existing private ride-hailing system. Their study highlights that not only the system but also travellers benefit from the shared rides. \cite{martinez2017assesing} provides profound insights into the environmental perspective, while \cite{zhang2021pool} analyses the travellers' perspective. \cite{fagnant2014travel} concludes that a single shared vehicle can substitute up to ten private cars.

The demand density and relative alignment of requests play a vital role in creating the pooling potential. In our method, demand of different size and structure may appear each day. Based on a topology of two German cities, \cite{zwick2022ride} analysed the impact of spatiotemporal demand distribution. \cite{shulika2024spatiotemporal} calculated KPIs for half a year of NYC trip requests data to uncover both the spatiotemporal and the density impact. 

Travellers are rational decision makers who choose the transportation mode that suits best their needs and is within the capabilities. \cite{gervzinivc2023potential} studies mode preference in urban areas and \cite{chavis2017development} in a general purpose flexible-route transit. For the specific case of ride-pooling perception, behavioural factors (such as a value-of-time) play a vital role. \cite{alonso2020value} studies preferences in the general ride-hailing and in \cite{alonso2021determinants} specifically for the ride-pooling case. \cite{lavieri2019modeling} focuses on the willingness to share a ride. \cite{krueger2016preferences} looks in depth at the value-of-time with a distinction between in and out of the vehicle. 
In \cite{bujak2024ride} we studied how, if explicitly revealed, individual diverse preferences would impact shareability.

The ride-pooling service is a complex problem. One needs to account for the spatiotemporal relations between travellers and find an optimal set of combinations. The first two seminar papers presenting a real-time applications were \cite{santi2014quantifying} (for a maximum of two travellers) and \cite{alonso2017demand} (no cap). \cite{bilali2020analytical} leverages shareability shadows to find feasible combinations. \cite{shah2020neural} proposes an algorithm with a matching based on a neural network. \cite{ke2021probability} works on the matching in an offline setting focusing on the two-sided aspect of the service. In our framework, we focus on the rationale that the shared ride must be more attractive than the alternatives, hence we build our base on the algorithm proposed by \cite{kucharski2020exmas}.

There are various methods of determining fares in a ride-pooling service. \cite{li2022pricing} and \cite{kucharski2020exmas} offer a sharing discount as a subject to a successful pairing. \cite{pandit2019pricing} compares fixed and dynamic (based on supply-demand) in a two-sided market. In the study by \cite{zhang2021pool}, researchers find a demand-supply equilibrium price. These studies are characterised by a single sharing discount level and an assumption of the population homogeneity. \cite{karaenke2023benefits} considers traveller's discomfort as a determinant for a fair fare in an ex-post pricing. \cite{fielbaum2022split} studies a cost-split based on the game-theory framework. \cite{jacob2021ride} introduces a two-sided queuing model and acknowledges a certain population heterogeneity by building two traveller classes. In \cite{bujak2024balancing} we proposed a pricing mechanism based on the general population distributions of behavioural traits.

\subsection{Study contribution}

We develop a ride-pooling pricing policy focused on a long-term performance of the service in a stochastic framework (non-deterministic travellers' decisions). 
Our framework assumes a heterogeneous population where we can distinct groups with specific value-of-time distributions. 
We consider travellers as utility-driven rational decision-makers who share rides only if they find them attractive (according to their individual behavioural traits). 
With received offers and previously taken shared rides, they build experience and become more or less likely to request a ride in the ride-pooling system. 

Our pricing policy integrates daily discount optimisation with the learning of individual behavioural traits. 
In the daily discount optimisation, we leverage both the short-term value of a ride (expected profit) and the attraction value (client loyalty - likelihood of joining the service). 
Based on our current knowledge of individual value-of-time, we optimise sharing discount to both attract travellers and improve the platform profit. 
In the learning process, we observe travellers' decisions (accepting or rejecting shared rides) and infer their behavioural class. 

Our study shows that:
\begin{enumerate}
    \item individual behavioural ride-pooling parameters can be learned from observing the sequence of offers made by platform and their acceptance by users.
    \item typically, it can be learned in less than $10$ days.
    \item we can introduce a single objective formula for platform operations, which maximises the expected long-term profit.
    \item by maximising this formula, the ride-pooling system becomes more efficient not only to the clients (increased satisfaction), but also to the operator (being more profitable) and to the system (by reducing the vehicle mileage).
\end{enumerate}

\section{Methodology}
We develop a framework for an assessment of a ride-pooling service in a stochastic setting and introduce a profit-maximising pricing policy (for the operator) within this framework\footnote{Python implementation available at \url{https://github.com/michallbujak/RidePoolingFrontiers/tree/main/Dynamic_pricing}.}. 
At a start of each day, travellers decide, based on their accumulated loyalty (satisfaction), whether they join the service.
The operator receives a batch of trip requests and aims to offer travellers rides (pooled or private).
Each traveller receives one offer: with whom he travels, with what discomfort (detour and delay compared to the private ride) and at what price.
The operator assigns travellers to rides (possibly shared) from the shareability set and sets a fare for each individual traveller. 
Travellers, based on their individual preferences (value-of-time), decide to accept or reject the offer. 
The operator wants the travellers to accept the offer, be satisfied with it (increased satisfaction) and, last but not least, make a profit. 
Every day, the operator observes how the clients accept or reject offered shared rides and has a chance to revise the strategy for the next day. 
Knowing the population distributions of latent variables (value of time) operator uses Bayesian inference to gain knowledge about individual travellers' attributes and satisfaction.

In Algorithm \ref{alg:overview}, we present the general overview of our framework. 
We work from two perspectives.
First, as the operator, we optimise and provide the service based on a limited (growing) knowledge.
Second, as a part of the framework, we mimic travellers' behaviour and sample their decisions according to the actual individual parameters.
\begin{algorithm}[!ht]
\caption{Algorithm overview}\label{alg:overview}
\begin{algorithmic}
    \Require \algopoints{Initial satisfaction}{Operator: assume the initial likelihood of clients joining the service (satisfaction).}
    \Require \algopoints{Population value-of-time}{Operator: learn the general population distribution of the value-of-time.}
    \For{day in Service Period}
        \State \algopoints{Day demand}{Framework: sample travellers decisions to join the service (according to their satisfaction).}
        \State \algopoints{Shareability set}{Operator: create a set of all potential shared rides with trip characteristics.}
        \State \algopoints{Fare optimisation}{Operator: optimise personalised fares (for rides in the shareability set).}
        \State \algopoints{Sample decisions}{Framework: sample travellers' decisions based on their actual behavioural class.}
        \State \algopoints{Update satisfaction}{Framework: update travellers' actual satisfaction.}
        \State \algopoints{Value-of-time Bayesian inference}{Operator: infer the value-of-time classes from travellers' choices.}
        \State \algopoints{Satisfaction estimation}{Operator: predicts current satisfaction of travellers. }
    \EndFor
    \State \algopoints{Performance Analysis}{Framework: analyse service performance: operator, travellers, system.}
\end{algorithmic}
\end{algorithm}

Our pricing policy within the framework can be divided into two reoccurring stages: day-to-day learning and within day optimisation. 
In the day-to-day learning, the operator uncovers individual classes (of the value-of-time) along with the accumulated service satisfaction. 
In the within day optimisation, based on the current knowledge of individual parameters, the operator provides clients service optimised for both short- and long-term profit.

In the within day process, the operator builds an offer (each traveller receives a single proposition of a private or a pooled ride).
\begin{itemize}
    \item First, the operator collects trip requests and, based on utility equations, constructs a shareability set - set of all potential shared rides (Sec. \ref{sec:utility_shareability}). 
    \item We step down to a ride level and conduct the personalised fare optimisation. We introduce the acceptance probability function, which conveys how likely a traveller is to accept a certain shared ride (Sec. \ref{sec:acceptance_prob}).
    \item The probability depends on the individual value-of-time (estimated by the operator) and the offered fare (control variable). 
    The actual outcome of the offer depends on the travellers' decisions - whether they accept or reject proposed shared rides, as detailed in Section \ref{sec:decision_realisation}. 
    \item To optimise the personalised fares, the operator looks at two aspects: current value of the ride and long-term attraction of travellers.
    The current value is calculated as the expected profit (Sec. \ref{sec:expected_profit}).
    \item The long-term attraction depends on the accumulated satisfaction of travellers (Sec. \ref{sec:satisfaction}) and is expressed as attraction value (Sec. \ref{sec:attraction_value}).
    \item Finally, we formulate the discount optimisation objective at a ride level and step up back to the system level where the operator conducts matching - creates an offer for clients (Sec. \ref{sec:objective_offer}). 
\end{itemize}

During the daily operations, travellers decide to accept or reject the proposed offer. 
While the discount optimisation is conducted in a stochastic setting, where accept/reject choices are assigned certain weights (probability), in the actual service realisation, the decision is binary. 
Hence, each day we sample the exact value-of-time realisation for all travellers and, by applying the utility equations, obtain decisions. 
The operator observes the decisions and infers the individual parameters. 
In Section \ref{sec:parameters_estimation} we describe the day-to-day learning process.
We discuss the Bayesian inference applied to learn the value-of-time in Section \ref{sec:vot_estimation}. 
Then, we define the update process for the predicted satisfaction in Section \ref{sec:predicted_satisfaction}.

\subsection{Utility and shareability set}\label{sec:utility_shareability}
In our framework, travellers are, by our assumption, rational decision-makers. 
They choose to accept a shared ride only if it is more appealing to them than the private alternative. 
To quantify the perceived experience (or offer), we introduce the utility. 

The utility is a measure of inconvenience experienced by a traveller. 
It depends on a trip characteristics (travel time, pick-up delay) and individual traits. 
Each traveller has a different utility for specific shared rides (different trip characteristics) a single value for a private ride (the shortest route, no pick-up delays). \citep{kucharski2020exmas}. 
How travellers perceive the experience (ride) is a subject to their individual perception.
We capture this behavioural heterogeneity by formulating the value-of-time as a variable. 
For traveller $i$ we define utility $U^{ns}_i$ and $U^s_{i, r_l}$ of a non-shared ride and shared ride $r_l$, respectively.
\begin{subequations}\label{eq:utilities}
\begin{align}
& U^{ns}_i = -\rho d_i - \beta_t t_i \label{eq:utility_ns}\\ 
& U^s_{i, r_l}(\lam{i}) = -(1 - \lam{i}) \rho d_i - \beta_t \beta_{s, k} (\hat{t}_i + \hat{t}^p_i), \label{eq:utility_s}
\end{align}
\end{subequations}
where the first terms refer to monetary cost ($\rho$ denotes fare ($\$/\mbox{km}$), $d_i$ requested trip length (km) and $\lam{i}$ the personalised sharing discount) and second terms to the travel time ($\beta_t$ represents value-of-time ($\$/\mbox{h}$), $\beta_{s, k}$ penalty for sharing ride with the $k-1$ co-travellers, $t_i$ private travel time, $\hat{t}_i$ travel time when pooled and $\hat{t}^p_i$ pick-up delay due to sharing). 
We say that a shared ride $r_l$ is attractive for traveller $i$ with a sharing discount $\lam{i}$ if and only if
\begin{equation}\label{eq:delta_utility}
    \Delta U_{i, r_l}(\lam{i}) = U^s_{i, r_l}(\lam{i}) - U^{ns}_i \geq 0.
\end{equation}

\subsubsection{Shareability set}\label{sec:shareability_set}
At the beginning of each day, the operator collects trip requests and, based on the utility, constructs a shareability set. 
The shareability set comprises all feasible shared rides, i.e. combinations of travellers along with their origins and destinations orders. 

The operator seeks a set of all feasible combinations of travellers (along with their origin and destination sequence) for a given demand.
To apply a hierarchical search method proposed by \cite{kucharski2020exmas}, we simplify assumptions and set a constant value-of-time and sharing discount. 
To ensure that all potential shared rides are captured, the operator takes the value of the most pooling-favourable group and proposes a high discount. 
As a result, he obtains a complete set of all potentially attractive shared rides.
The operator, when constructing the offer to the travellers, will only consider rides from the shareability set (with optimised personalised fares).

\subsection{Acceptance probability}\label{sec:acceptance_prob}
Travellers experience longer travel time when choosing a shared ride, yet they pay a lower fare. 
We measure the relation between the two opposing factors via the utility equations. 
In this section, we step down to a ride level and analyse how likely a traveller is to accept a shared ride at a certain fare (personalised sharing discount).
For each ride in the shareability set, the operator constructs the individual (for each traveller) acceptance probability function (of the discount).

Utility equations (eq. \ref{eq:utilities}) lead to the utility difference in eq. \ref{eq:delta_utility}. 
In our setting, the value-of-time is a random variable, hence the $\Delta U_{i, r_l}(\lam{i})$ is also a random variable. 
The decision distribution whether a traveller accepts a shared ride depends not only on the sharing discount but also on the distribution of the $\beta_t$ (value-of-time).
Hence, we introduce an explicit notation $\Delta U_{i, r_l}(\beta_t, \lam{i})$.
We can transform the utility equations and obtain the following condition. 

\begin{equation}\label{eq:beta_t_reformulation}
    \Delta U_{i, r_l} (\beta_t, \lam{i}) \geq 0 \iff \lam{i} \rho d_r - \beta_t (\beta_s t^s - t^n) \geq 0 \iff \frac{\lam{i} \rho d_r}{\beta_s t^s - t^n} \geq \beta_t.
\end{equation}

Let $F_{\beta_t}$ be a cumulative distribution function of the $\beta_t$, then

\begin{equation}\label{eq:beta_t_cdf}
    \prob (\Delta U_{i, r_l} (\beta_t, \lam{i}) > 0) = F_{\beta_t} (\frac{\lam{i} \rho d_r}{\beta_s t^s - t^n}).
\end{equation}

We should note that each traveller has his value-of-time distribution, hence knows the actual probability that he will accept the offered shared ride (at the proposed fare). 
However, the operator does not know the actual individual value-of-time but relies on estimates. 
Hence, the acceptance probability function calculated by the operator is an approximation, not the actual probability.

\subsection{Decision-driven realisation of a shared ride}\label{sec:decision_realisation}
Travellers decide to accept or reject an offered shared ride according to the proposed sharing discount and their individual value-of-time (eq. \ref{eq:utilities}).
The operator, who aims to find optimal personalised sharing discounts, considers possible outcomes of the offer. 
Satisfied clients accept the offered shared ride while unsatisfied reject and are served with the private rides.
In this section, we describe how the accept/rejection decisions shape the final realisation of the service.

\begin{figure}[!ht]
    \centering
    \includegraphics[width=0.9\linewidth]{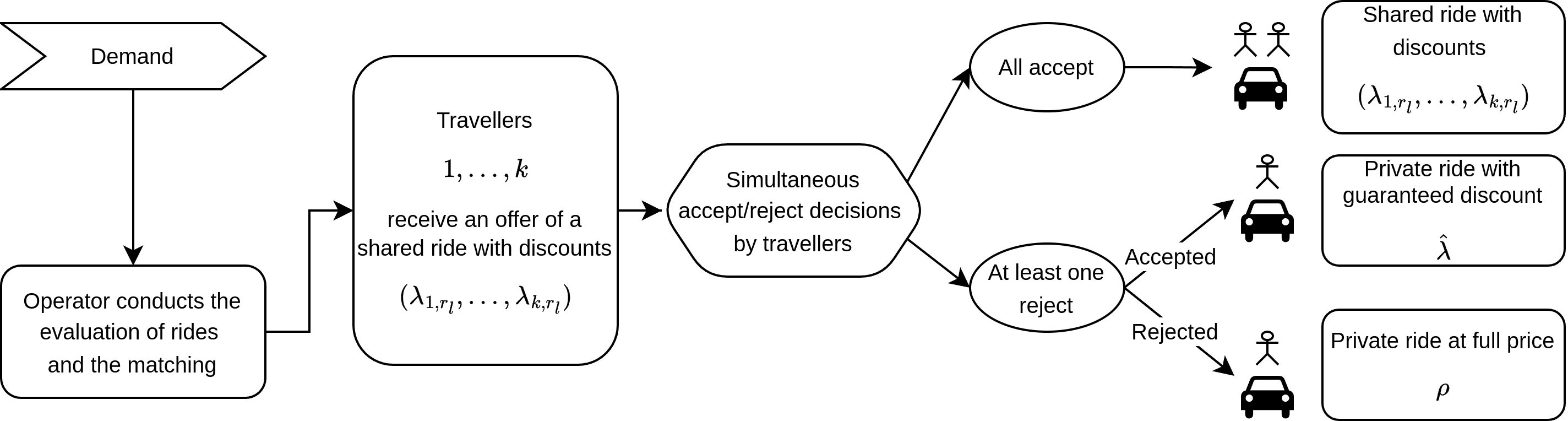}
    \caption{Travellers who are assigned a shared ride independently decide to accept or reject the offer.}
    \label{fig:pricing_scenarios}
\end{figure}

During the fare optimisation process, the operator considers all rides in the shareability set. 
For every shared rides, he fits personalised discounts, which are at least at the level of the guaranteed discount. 
By applying techniques described in Section \ref{sec:acceptance_prob}, the operator can assess the probability that a ride will be accepted at a certain fares (based on his estimates of client's value-of-time distribution). 
We assume that, once a shared ride is offered, travellers reach their accept/reject decisions independently. 
If all co-travellers accept the ride, they receive sharing discounts according to the offer.
However, if the ride is rejected by at least one, all co-travellers are served with private rides. 
Those travellers who accepted the ride receive a guaranteed discount.
Travellers who rejected the shared ride pay the full fare.
The pricing scheme is depicted in Figure \ref{fig:pricing_scenarios}.

Certain travellers who do not align well for pooling are immediately offered private rides.
Those travellers receive a guaranteed discount (motivated by real-world applications \cite{uberpool}). 
We assume travellers always accept a private ride.

\subsection{Expected Profit}\label{sec:expected_profit}
In previous sections, we described the individual acceptance probability (Sec. \ref{sec:acceptance_prob}) and the pricing cases (Sec. \ref{sec:decision_realisation}). 
Here, we introduce the expected profit measure, which is our proxy to a ride's value.

Travellers, who are offered a shared ride, decide whether they accept it or reject. 
As a result, for each shared ride comprising $k$ travellers, we have $k$ independent decisions.
Each of those decisions has a certain probability. 
As a result, we have $2^k$ configurations of decisions with the corresponding probability. 

We introduce the expected profit which incorporates all decision configurations.
The proposed measure combines the following three factors: revenue, mileage and flat fleet upkeep cost. 
Revenue is strictly calculated as a requested trip length multiplied by a fare (sum for shared rides). 
Mileage serves as the operations cost of a vehicle (e.g. fuel). 
Flat fleet upkeep corresponds to the cost of the readiness of a car (in case of non-CAVs, also salary for a driver).
Let $\betrl = (\beta_{t, 1}, \ldots, \beta_{t, k})$ denote the estimated value-of-time distribution of the respective travellers in $r_l$.
The expected profit $\Gamma$ for a ride $r_l$ with discount vector $\lambda^{r_l} = (\lambdas)$ is calculated as
\begin{equation}\label{eq:expected_profitability}
    \Gamma^{r_l}(\lambda, \betrl) = R^{r_l}(\lambda, \betrl) - \zeta \gamma^{r_l}(\lambda, \betrl) - C*\theta^{r_l}(\lambda^{r_l}, \betrl),
\end{equation}
where $R(\cdot)$ denotes the expected revenue, $\zeta$ - mileage sensitivity, $\gamma(\cdot)$ - expected mileage, $C$ - flat upkeep cost, $\theta(\cdot)$ - the expected number of required vehicles.

By our assumption, any private ride is accepted.
Hence, for traveller $i$'s private ride we write subscript $i$ instead of superscript $r_l$ (e.g. $\Gamma_i$ instead of $\Gamma^{r_l}$, $R_i$ instead of $R^{r_l}$) and omit the second argument, i.e. $\beta_t$ distribution.
The private revenue $R_i$ and mileage $\gamma_i$ are calculated as:
\begin{subequations}\label{eq:ex_rev_ns}
    \begin{align}
        & R_i(\gd) = \rho (1 - \gd) d_i, \\
        & \gamma_i = d_i,
    \end{align}
\end{subequations}
where $\gd$ denotes the guaranteed discount and $d_i$ request length (symbols follow notation from Section \ref{sec:utility_shareability}).
Therefore, the expected profit of the private ride is
\begin{equation}\label{eq:ex_profitability_ns}
    \Gamma_i(\gd) = \rho (1 - \gd) d_i - \zeta d_i - C.
\end{equation}

To calculate the expected revenue of a shared ride, we must devise decisions vectors (Figure \ref{fig:pricing_scenarios}). 
We have a single case where all travellers accept and the ride is effectively shared. 
Then, we have all combinations where at least one traveller rejects the ride.
Let $X_i(\lam{i}, \beti)$ be a binary indicator representing whether traveller $i$ accepts the shared ride $r_l$ with the sharing discount $\lam{i}$ when his value-of-time distribution follows $\beti$.
For the notation conciseness, let $\betrl = (\beta_{t, 1}, \ldots, \beta_{t, k})$ denote distribution of the value-of-time of travellers in $r_l$ and $\lambda^{r_l}=(\lambdas)$ offered discounts.
The weighted expected revenue for the case when the shared ride is accepted is:
\begin{equation}\label{eq:ex_rev_sh_acc}
    R^{r_l}_a(\lambda^{r_l}, \betrl) = \prob(\Pi_{1 \leq i \leq k} X_{i, r_l}(\lam{i}, \beti) = 1) \left(\sum_{1 \leq i \leq k} R_i(\lam{i}) \right),
\end{equation}
where $R_i(\lam{i}) = \rho(1-\lam{i})d_i$ (as in eq. \ref{eq:ex_rev_ns}). 
For the part where the shared ride is not realised (at least one rejection), travellers who accepted get the guaranteed discount ($R_i(\gd) = \rho (1 - \gd) d_i$) and those who rejected pay the full fare ($R_i(0) = \rho d_i$).
Therefore, the weighted expected revenue for the rejected ride $R_{r_l}^r(\lambda^{r_l}, \betrl)$ is 
\begin{equation}\label{eq:ex_rev_sh_rej}
    \sum_{0 \leq s < k} \sum_{\substack{i_1 + \ldots + i_k = s, \\ i_j \in \{0, 1\}}} \prob(X_{1, r_l}(\lam{1}, \beta_{t, 1}) = i_1, \ldots, X_{k, r_l}(\lam{i}, \beta_{t, k}) = i_k) \left(\sum_{1 \leq j \leq k} i_j * R_j (\gd) + |1 - i_j| * R_j(0)\right).
\end{equation}
We calculate the total expected revenue $R^{r_l}(\lambda^{r_l})$ of the shared ride $r_l$ as
\begin{equation}\label{eq:ex_rev_sh}
    R^{r_l}(\lambda^{r_l}, \betrl) = R_a^{r_l}(\lambda^{r_l}, \betrl) + R_r^{r_l}(\lambda^{r_l}, \betrl).
\end{equation}

To determine the expected distance, we devise scenarios analogically to eq. \ref{eq:ex_rev_sh_acc} and eq. \ref{eq:ex_rev_sh_rej}. 
However, the procedure is simplified, because when the shared ride is not realised, the distance is the same irrespective of the exact rejection distribution.
We calculate the expected distance (vehicle mileage) as:
\begin{equation}\label{eq:ex_veh_dist}
    \gamma_{i, r_l}(\lambda^{r_l}, \betrl) = \prob(\Pi_{1 \leq i \leq k} X_{i, r_l}(\lam{i}, \beti) = 1) d_{r_l}^s + (1 -  \prob(\Pi_{1 \leq i \leq k} X_{i, r_l}(\lam{i}, \beti) = 1))(\sum_{1 \leq t \leq k} d_t),
\end{equation}
where $d_{r_l}^s$ denotes distance of the realised shared ride $r_l$ and $d_i$ represents individual distance of a private ride for traveller $i$.
The last element required to measure the expected profit according to the eq. \ref{eq:expected_profitability} is the expected number of vehicles. 
It is calculated as
\begin{equation}
    \theta^{r_l}(\lambda^{r_l}, \betrl) = \prob(\Pi_{1 \leq i \leq k} X_{i, r_l}(\lam{i}, \beti) = 1) + k(1-\prob(\Pi_{1 \leq i \leq k} X_{i, r_l}(\lam{i}, \beti) = 1)).
\end{equation}

\subsection{Satisfaction}\label{sec:satisfaction}
To understand and control the long-term performance of the ride-pooling service, we must consider how clients' experience shapes the future demand. 
Satisfied travellers are more likely to join the pooling service again, whereas dissatisfied ones will avoid pooling in the future. 
To measure travellers' attitude towards the service, we introduce satisfaction. 
The satisfaction conveys how travellers perceive the experience and determines the future demand (travellers opt to join or drop out).

To measure a current satisfaction of a traveller, we consider his initial level and experience gathered with the service. 
Specifically, the experience is measured via utility gain compared to a private ride ($\Delta U$ in eq. \ref{eq:delta_utility}).
There are three scenarios for a traveller: (i) joins the service and experiences a shared ride; (ii) joins the service and rejects a shared ride; (iii) either joins and is served with a private ride (if offered a shared ride, he accepted but the ride was not realised) or he does not join the service. 
In the first case (i), the traveller recognised the shared ride as attractive and so do his co-travellers. 
He experiences the attractive shared ride ($\Delta U \geq 0)$) and increases his satisfaction. 
In the second case (ii), the traveller considers the offered ride unattractive ($\Delta U \leq 0$). 
Therefore, he rejects it and is served with a private ride at a full fare. 
The level of how unattractive the shared ride offered to him defines by how much his satisfaction decreases. 
The third case (iii) appears under several conditions: the traveller does not join the service; or is offered a private ride; or is offered an attractive shared ride which is rejected by one of his co-travellers. 
In any of those cases, the traveller does not experience shared rides.

An offered shared ride combines the trip characteristics with the personalised fare. 
In our simulation process, the traveller must make a decision to either accept or reject the offer. 
We sample his exact value-of-time realisation according to his actual distribution (latent to the operator) and obtain the corresponding $\Delta U$ value.

Formally, let traveller $i$ belong to a class with $\beti$ distribution of the value-of-time. 
At a start, he has a satisfaction $s_{i, 0}$.
We update the satisfactions after each day as follows.
Let $J(i, n)$ indicate whether traveller $i$ joins the service on day $n$ and is offered a shared ride (denoted $r_l$).
Let $Z(i, r_l)$ represents whether the shared ride $r_l$ is accepted by $i$'s co-travellers (a binary indicator).
\begin{equation}\label{eq:satisfaction}
    s_{i, n+1} =
    \begin{cases}
        s_{i, n} + \Delta U_{i, r_l}(\lam{i}, \beti), \mbox{ if (i), i.e. } J(i, n) = 1, \Delta U_{i, r_l}(\lam{i})\geq 0, Z(i, r_l)=1, \\
        s_{i, n} + \Delta U_{i, r_l}(\lam{i}, \beti), \mbox{ if (ii), i.e. } J(i, n) = 1, \Delta U_{i, r_l}(\lam{i})< 0,  \\
        s_{i, n}, \mbox{ otherwise (iii).}
    \end{cases}
\end{equation}
In (i) the satisfaction increases, in (ii) decreases, and in (iii) does not change.
In the simulation framework, on day $n$, for the traveller $i$ who is offered shared ride $r_l$, we sample a realisation of his value-of-time $b_{t, i}$ according to $\beti$. 
Then, we deterministically calculate $\Delta U_{i, r_l}(\lam{i}, b_{t, i})$.
From here, we obtain the $s_{i, n+1}$.

The probability that a traveller $i$ joins service on day $n$ depends on the current $s_{i, n}$. 
We adopt the \textit{s-shaped} learning curve as proposed in \cite{ghasemi2025momas} based on the sigmoid function.
Namely, probability $\prob(J(i, n)=1)$ that the traveller $i$ joins the service on day $n$
is calculated as:
\begin{equation}\label{eq:participation_prob}
    \prob(J_{i, n}=1) = \frac{1}{1 + \exp{(-s_{i, n})}}.
\end{equation}
The s-shaped curve puts emphasis the fact that new experience has the strongest impact on travellers who are neutral. 
Those who lean towards either side are less likely to change their perception drastically. 

We assume the initial individual satisfaction levels are known to the operator, who can therefore approximate their current level.

\subsection{Attraction value}\label{sec:attraction_value}
The operator should consider ramifications of his offer on the future demand. 
As detailed in Section \ref{sec:satisfaction}, clients, based on their experience, change attitude towards pooling and become more or less likely to join the service in the future. 
Because the ride-pooling heavily depends on the compatibility of requests, reaching higher demand is crucial to find well-aligned shared rides. 
Yet, certain trips have higher priority for the operator (long trips, well aligned) than others. To quantify how important is to encourage a traveller to rejoin the service, we introduce the attraction value.

To propose a computationally feasible solution, we take into regard two components: reappearance of the shared ride (for which we optimise discounts) and the private rides offered to all co-travellers. 
The two scenarios are weighted according to their likelihood.
We should note that, for the attraction value, we associate a traveller with a trip request. While questionable in practice, one could argue that certain services are used daily and requests likely appear repeatedly.

Let the shared ride $r_l$ comprise travellers $1, \ldots, k$ and $s_{i, n}$ represent satisfaction of traveller $i$ on day $n$.
The operator estimates the value-of-time distribution of traveller $i$ as $\beti$.
For a considered discount levels $\lambda^{r_l} = (\lambdas)$, the operator calculates potential utility changes $\Delta U_{i, r_l}(\lam{i})$ according to eq. \ref{eq:delta_utility}.
To calculate the (potential) future satisfaction, the operator instead of variable $\Delta U_{i, r_l}(\lam{i})$ applies $E(\Delta U_{i, r_l}(\lam{i}))$ to eq. \ref{eq:satisfaction} and derives $s_{n, i+1}(\lam{i}, \beti)$ as a value.
The increment in the probability that traveller $i$ who is offered ride $r_l$ with a discount $\lam{i}$ is
\begin{equation}\label{eq:prob_increase}
    \Delta p_{i, n}(\lam{i}, \beti) = \frac{1}{1+\exp{(-s_{i, n+1}(\lam{i}, \beti))}} - \frac{1}{1+\exp{(-s_{i, n})}}.
\end{equation}
The operator estimated what is the change in the likelihood of a traveller $i$ reappearing. 
Now, to calculate the attraction value, he incorporates the information on how valuable $r_l$ is based on the expected profit $\Gamma^{r_l}(\lambda^{r_l})$ and $\Gamma_i$ introduced in Section \ref{sec:expected_profit}.

The attraction value $F_{r_l}$ of the shared ride $r_l$ has two components: shared $F^s_{r_l}$ and private $F^{p}_{r_l}$.
The shared part $F^s_{r_l}$ represents the scenario where all travellers reappear in the system and is calculated as follows.
\begin{equation}\label{eq:future_s}
    F^s_{r_l}(\lambda^{r_l}, \beta_t^{r_l}) = \Pi_{i \leq k} \left( \Delta p_{i, n}(\lam{i}, \beti) \right) \Gamma^{r_l}(\lambda^{r_l}).
\end{equation}
The private component corresponds to a scenario, where at least one of the travellers does not reappear in the system.
It is calculated as:
\begin{equation}\label{eq:future_ns}
    F^{p}_{r_l}(\lambda^{r_l}) = \sum_{i \leq k} \Delta p_{i, n}(\lam{i}) \Gamma_i \left(1-  \Pi_{j \leq k, j \neq i}(\prob(X_{j, r_l}(\lam{j})=1) \right).
\end{equation}
Finally, the attraction value $F_{r_l}$ for ride $r_l$ with the proposed discounts $\lambda^{r_l}$ is defined as:
\begin{equation}\label{eq:attraction_value}
    F_{r_l}(\lambda^{r_l}) = F^{s}_{r_l}(\lambda^{r_l}) + F^{p}_{r_l}(\lambda^{r_l}).
\end{equation}

\subsection{Objective maximisation and offer}\label{sec:objective_offer}
To find the optimal offer, the operator works at two levels. 
First, at a ride level, he optimises fares to maximise the local objective function (sum of the expected profit and the attraction value). 
Second, once the optimal fares are found, he conducts a matching to select rides for the offer. 
In the offer, each traveller should receive a single proposition of a shared or private ride. 
To construct the offer, the operator maximises a global objective function (sum of the values of the local objective functions on selected rides). 
In this section, we detail the local and global objectives along with tools required for the matching.

First, let us consider the discount optimisation at a ride level. The operator seeks a local objective function that guarantees both short- and long-term profitability.
The short-term profitability guarantees that the operator achieves a high profit on a current service day.
The long-term profitability is focused around building a loyal client base and therefore increasing the demand levels.
We address the first point with the expected profit (Section \ref{sec:expected_profit}) and the second with the attraction value (Section \ref{sec:attraction_value}).
We emphasise a fact that the operator controls the fare levels $\lambdarl$, but the objective functions are also determined by estimates of the individual value-of-time $\betrl$.
Combining the short- and long-term perspectives, for a shared ride $r_l$ the local objective function $\Upsilon(\lambdarl, \betrl)$ (the total expected profit) is introduced as
\begin{equation}\label{eq:obj_func}
    \Upsilon_{r_l}(\lambdarl, \betrl) = \Gamma^{r_l}(\lambdarl, \betrl) + \beta_F F_{r_l}(\lambdarl, \betrl),
\end{equation}
where $\beta_F$ is the attraction value sensitivity.
We seek a discount vector $\lambda_{r_l}^* = (\lam{1}^*, \ldots, \lam{k}^*)$ that:
\begin{equation}\label{eq:obj_eq}
    \lambda_{r_l}^* = \underset{\lambdarl\in [\gd, \nu]^k}{\mbox{argmax}} \Upsilon_{r_l}(\lambdarl, \betrl),
\end{equation}
where $[\gd, \nu]$ is an interval between the guaranteed discount $\gd$ and the maximal discount $\nu$ ($\nu < 1$).
Once we find the desired vector $\lambda_{r_l}^*$, each ride has its maximal value
\begin{equation}\label{eq:max_value}
    M(r_l) = \upsilon(\lambda_{r_l}^*).
\end{equation}
At this stage, we found optimal discounts for each ride individually in our shareability set.

The second stage is at the system level. The operator optimised each ride and assigned them maximal values. 
Now, the goal is to find a set of rides (subset of the shareability graph) such that each traveller is served exactly once and the global objective is maximised.
As the global objective we propose the sum of the maximal values of the selected rides ($M(r_l)$).

Formally, let $\mathcal{S}$ denote the shareability set, $\mathcal{T}$ the set of all travellers, and $T(r_l)$ set of travellers who comprise ride $r_l$.
For the selected set of rides $S$ we define the global objective $\Psi(S) = \sum_{r_l \in S} M(s)$. 
We seek an assignment $A: \mathcal{T} \xrightarrow{} \mathcal{S}$ that satisfies:
\begin{itemize}
    \item $A$ is a function - each travellers is offered a single (private or shared) ride.
    \item $\forall_{t \in \mathcal{T}} \forall_{s \in T(A(t))} A(t) = A(s)$ - if any traveller is assigned to a ride, all traveller who comprise this ride are assigned to it.
    \item $\forall_{B: \mathcal{T} \xrightarrow{} \mathcal{S}} \Psi(A(\mathcal{T})) \geq \Psi(B(\mathcal{T}))$ - the assignment maximises the global objective. 
\end{itemize}
To find the assignment $A$, we adopt the linear integer programming solution provided with PuLP Python open-source library \cite{mitchell2011pulp}.

We should note that in our calculations, we discretize value-of-time and therefore also explore a discrete finite space of discounts. 
Our process finds the optimal solution for the system by optimising discounts at a ride level, when our objective is the sum of the local objectives (as we introduced).
The formal proof and details are presented in \cite{bujak2024balancing}.

\subsection{Parameters estimation}\label{sec:parameters_estimation}
Preceding sections were dedicated to within day fare optimisation.
Here, we focus on the day-to-day learning process. 
With each day of operations, the operator learns the individual distributions of the value-of-time and adapts the pricing policy accordingly. 

The operator not only provides the service but also collects information by observing travellers' choices.
Thanks to the increasing amount of information, he creates an offer which is both more profitable and more appealing to the travellers. 
To achieve such goals, the operator needs to understand individual parameters and uncover the decision-making process of specific travellers.
Travellers who rejected a shared ride are more likely to belong to a high value-of-time class and we know that their satisfaction decreased. 
Ones who experienced a shared ride found the offer appealing and increased their satisfaction. 
The operator cannot infer any new information when a traveller decided not to join the service or travelled by a private ride. 
Therefore, the learning process occurs only for travellers who were offered a shared ride.

\subsubsection{Value-of-time estimation}\label{sec:vot_estimation}
According to stated preference studies (\cite{alonso2021determinants}, \cite{zwick2022ride}), travellers belong to one of (a finite number of) groups with a distinct value-of-time distribution.
To determine the individual class membership, we propose the Bayesian inference.
With each observation, the operator updates the class distribution probabilities. 

Let the value-of-time classes be $C_1, \ldots, C_k$. 
From the operator's perspective, the prior probability (at the start of day $n$) that traveller $i$ belongs to class $C_i$ is $\prob_a(i \in C_i)$.
On day $n$, traveller $i$ is offered a shared ride $r_l$ with a discount $\lam{i}$.
To mimic the actual behaviour (knowing his actual class $C_j$), we sample the decision from $X_{i, r_l}(\lam{i}, \beti)|i\in C_j$ and denote it $d_{i, r_l}$ (accept or reject).
The operator, unaware of the actual classes, based on the observed choices, calculates the posterior (at the end of day $n$) probability that the traveller $i$ belongs to a class $C_j$ as:
\begin{equation}\label{eq:vot_estimate}
    \prob(i\in C_j|X_{i, r_l}(\lam{i})=d) = \frac{\prob_a(X_{i, r_l}(\lam{i})=d|i\in C_j) \prob_a(i \in C_j)}{\prob_a(X_{i, r_l}(\lam{i})=d)}.
\end{equation}
The operator knows distribution within each class $C_j$ (that is our assumption), hence $\prob_a(X_{i, r_l}(\lam{i})=d|i\in C_j)$ is explicitly calculated from the distribution of $C_j$. 
$\prob_a(i \in C_j)$ is the explicit prior distribution and $\prob_a(X_{i, r_l}(\lam{i})=d)$ is calculated as $\sum_{s\leq k} \prob(X_{i, r_l}(\lam{i})=d|i\in C_s)$.

We can measure the class estimation error as follows.
Let the operator's predicted probability that the traveller $i$ belongs to class $C_j$ be $\phi(i, j)$ and $\Phi(i)$ denote the actual class of the traveller.
The class estimation error for traveller $i$ is
\begin{equation}\label{eq:vot_est_err}
    1 - \phi(i, \Phi(i)).
\end{equation}

\subsubsection{Predicted satisfaction}\label{sec:predicted_satisfaction}
The satisfaction determines whether a traveller joins the service and hence drives the attraction value. 
The true satisfactions is known only to travellers themselves, but the operator requires estimates to assess the participation probability change (and therefore the attraction value).
The satisfaction (both actual and estimated) is updated if a traveller effectively shares a ride (experiences the benefits) or rejects a ride (instead of an unattractive shared offer served with a private one at a full fare).

To deterministically calculate the actual satisfaction change, each day, we sample a realisation from the actual class of a traveller.
When the value-of-time is fixed, we can explicitly calculate their utility and add to the previous satisfaction.

The operator does not know the actual satisfaction, so he relies on estimates. 
The estimates are updated once the decisions are collected and the class membership probability is recalculated.
Let $\prob(i \in C_j)$ denote probability that traveller $i$ is in class $C_j$ (with a known to the operator distribution of the value-of-time) and $s_{i, n-1}$ represent previous estimated satisfaction.
The current satisfaction for traveller $i$ offered ride $r_l$ is estimated as:
\begin{equation}\label{eq:satisfaction_update}
    s_{i, n} = s_{i, n-1} + \sum_l \prob(i \in C_l)  E(\Delta U_{i, r_l} (\lambda)|i \in C_l).
\end{equation}

\section{Results}
In this section, we present numerical results with a case study for our pricing policy within the proposed framework. 
We experimentally show that the policy achieves outstanding results for a long-term ride-pooling service. 

First, in Section \ref{sec:experiment_design}, we describe the experiment design.
In Section \ref{sec:res_profit}, we scrutinise the expected and actual profit. 
We study travellers' perspective -- satisfaction and participation probability (Section \ref{sec:res_satisfaction}), and follow with the system properties -- sharing and acceptance rates (Section \ref{sec:res_sharing_performance}).
Next, we look at the discount distribution changes with each day (Section. \ref{sec:res_fitted_discounts}). 
Then, in Section \ref{sec:res_class_estimation}, we analyse the learning process.
We conclude by comparing our pricing policy to a flat-pricing baseline and show the impact of the learning process and the attraction value on our personalised policy (Section \ref{sec:res_baseline}).

\subsection{Experiment design}\label{sec:experiment_design}
We showcase our method, using as close to realistic assumptions as we can. 
We set the experiment in Manhattan, NYC. 
Our goal is to mimic daily operations, where travellers consider ride-pooling as their potential mobility service for a regular route (e.g. on the way from work to home). 
We use the open-source data publicly available at \cite{NYCdata} and sample a medium-density (\cite{shulika2024spatiotemporal}) batch of $300$ requests from 2024-01-08 at 16:00 to 16:30 (peak hours on the first non-holiday Monday of the year).
It is a pool of all potential clients who, we assume, all start with null satisfaction ($s_{i, 0}= 0$). 
We monitor results during $20$ days of operations, where we assume the (maximum) demand is repeated daily.

We set the guaranteed discount at $\gd = 5\%$ (following \cite{uberpool}) and a maximum discount of $\nu =40\%$ (flat for the shareability set generation). 
For the penalty for sharing, we take arbitrary levels of $\beta_{s, 2} = 1.148, \beta_{s, 3}= 1.4$ and $\beta_{s, 4} = 2$ (for $2, 3$ and $4$ co-travellers). 
To realistically reflect population heterogeneity of the value-of-time, we follow results by \cite{alonso2021determinants} as adopted in \cite{bujak2024ride} summarised in Table \ref{tab:vot_values}.
\begin{table}[!ht]
    \centering
    \resizebox{0.7\textwidth}{!}{
    \begin{tabular}{cc|c||c|c|c|c}
          & &  & $C_1$ & $C_2$ & $C_3$ & $C_4$ \\ \hline         
        \multirow{2}{*}{Value-of-time} & \multirow{2}{*}{$\beta_t$} & Mean & 16.98 & 14.02 & 26.25 & 7.78 \\ 
        & & Std.dev. & 0.318 & 0.201 & 5.777 & 1 \\ \hline
        Share & $\prob(i\in C_j)$ & & 29\% & 28\% & 24\% & 19\% \\ \hline
    \end{tabular}}
    \caption{\small The value-of-time distribution of population in four classes $C_1, C_2, C_3$ and $C_4$.}
    \label{tab:vot_values}
\end{table}
We set the attraction value sensitivity $\beta_F=1$ (eq. \ref{eq:obj_func}).
The reminder of parameters follows the default ExMAS values \citep{kucharski2020exmas}.

\subsection{Profit}\label{sec:res_profit}
The primary goal of the proposed pricing policy is to maximise the expected profit (also by attracting higher demand).
Here, we show that the profit grows and becomes less variant with time.

\begin{figure}[!ht]
    \centering
    \includegraphics[width=0.8\linewidth]{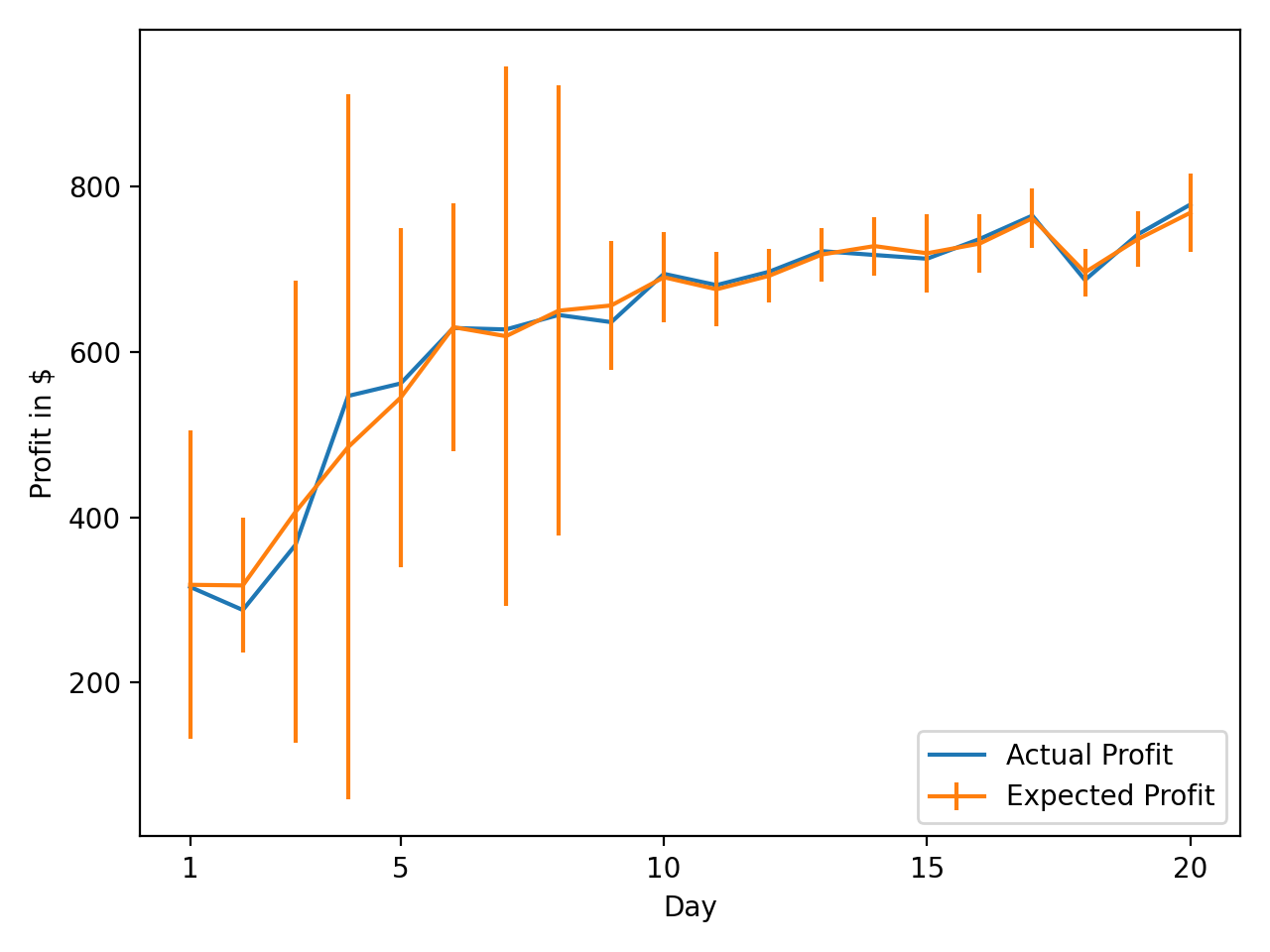}
    \caption{Profit of the system. In orange the predicted mean and variance, in blue actual realisation. On $x$-axis day, on $y$-axis profit in the monetary unit ($\$ $). Both actual and expected profit increases with time and variance of predictions decreases.}
    \label{fig:service_performance}
\end{figure}

Our optimisation process leverages distributions of value-of-time (hence, the accept/reject decisions with corresponding probabilities), yet the final performance is driven by a single realisation. 
To analyse how accurate and stable our results are, for each day we calculate the expected value and the theoretical variance.
To calculate the theoretical values, we consider actual distributions of individual value-of-time and resulting acceptance/rejection decisions, along with their respective probability.
It translates to various realisations where each has a certain probability.
In Figure \ref{fig:service_performance}, we show the theoretical values and a single realisation (the actual data contains $20$ points, but we plot it as continuous for the presentation clarity). 
The actual (realised) values are close to the theoretical mean and well within the range of the variance. 
The greatest variance appears when the operator has already attracted more travellers (positive average satisfaction), and yet he is unsure of their value-of-time. 
Later, the individual value-of-time is uncovered and the ratio of variance to mean becomes minuscule.

\subsection{Satisfaction}\label{sec:res_satisfaction}
Travellers who appreciate offered shared rides increase their satisfaction and more likely join the service in the following days. 
In our model, the actual satisfaction is known only to travellers themselves. 
The operator relies on estimates. 
Here, we compare the satisfaction estimates to the actual values and the resulting participation probability.

\begin{figure}[!ht]
    \centering
    \includegraphics[width=0.8\linewidth]{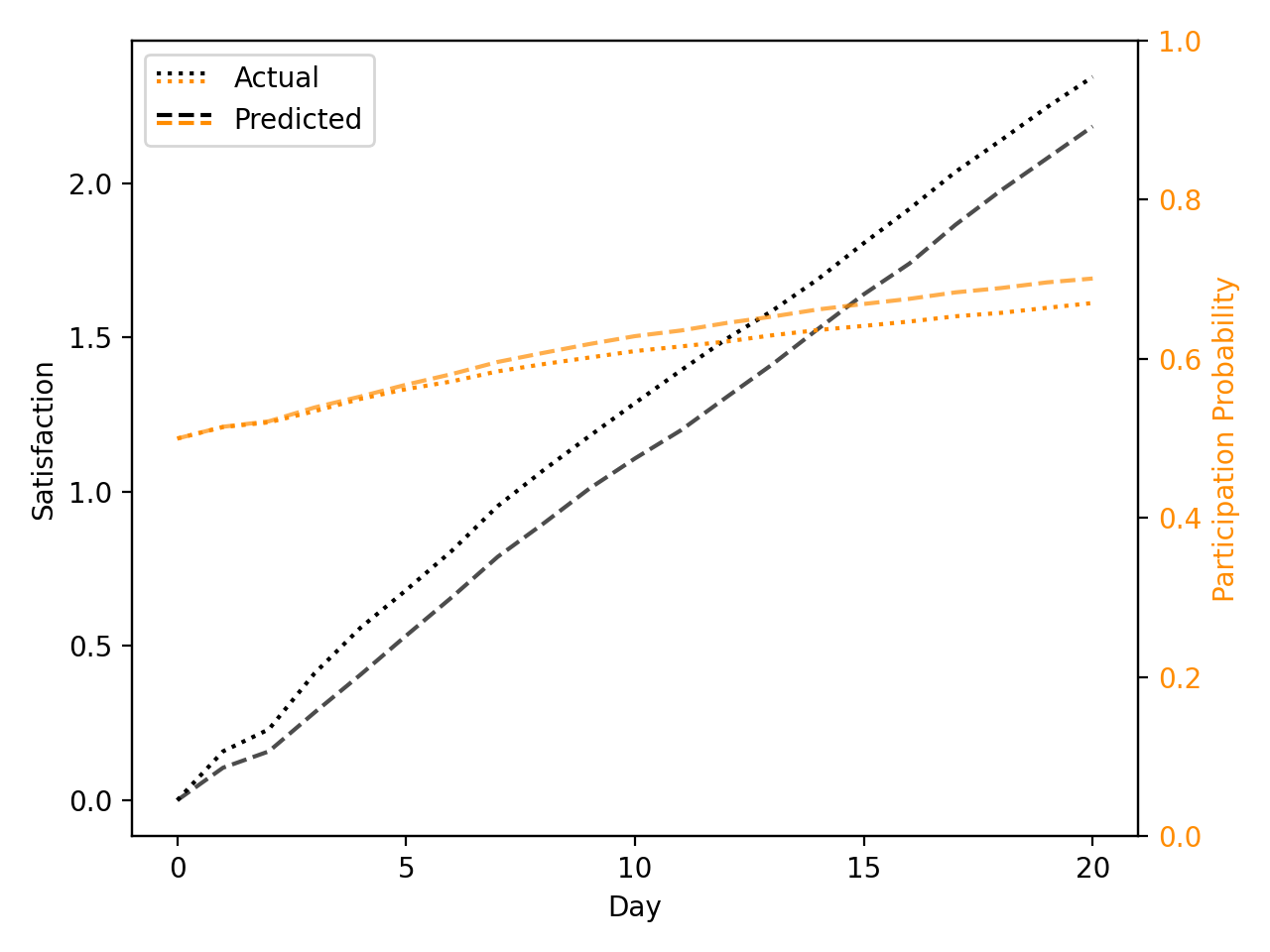}
    \caption{The mean satisfaction of travellers (black) and the mean participation probability (orange). Dotted represents the actual (latent to the operator) values and dashed denotes the operator's estimates. The estimates are near the actual values and replicate trends closely.}
    \label{fig:satisfaction}
\end{figure}

Everyone starts, by our assumption, with $0$ satisfaction ($50\%$ participation probability) \citep{ghasemi2025momas}. 
In Figure \ref{fig:satisfaction}, we present the change of the mean satisfaction with the service period.
The operator's projections are close to the actual accumulated satisfaction. 
Bayesian inference supports quick convergence of the class membership distribution and, therefore, the satisfaction estimation. 
Yet, the initial discrepancy between estimates and the actual values leads to a long-term disparity, as the early difference between two values is not mitigated with time and the shift remains. 
We should note that only travellers who are offered shared rides update their experience.
Certain travellers remain always incompatible for pooled rides and therefore remain at the zero satisfaction level. 
However, those who are offered shared rides, on average, improve their satisfaction. 
As a result, the demand density increases, which further endorses the system performance.

We note that satisfaction is not linearly correlated to participation probability (eq. \ref{eq:participation_prob}). 
Hence, even though the mean estimated participation probability is higher than the actual, in this case, the mean estimated satisfaction is lower than the actual. 
As an example, when two people both increase their satisfaction from $2$ to $3$, the mean participation probability increases by $7.1\%$; while when one person remains at $2$ satisfaction and second increases from $2$ to $5$, the participation probability increases by $5.6\%$ on average.

\subsection{Sharing performance}\label{sec:res_sharing_performance}
Shared rides are appreciated by the system (city, environment) thanks to the mileage and vehicle fleet size reduction, less parking space required, etc. (\cite{martinez2017assesing}, \cite{engelhardt2019quantifying}).
In this section, we look at how successful is our pricing strategy in promoting the shared mobility.

\begin{figure}[!ht]
    \centering
    \includegraphics[width=0.8\linewidth]{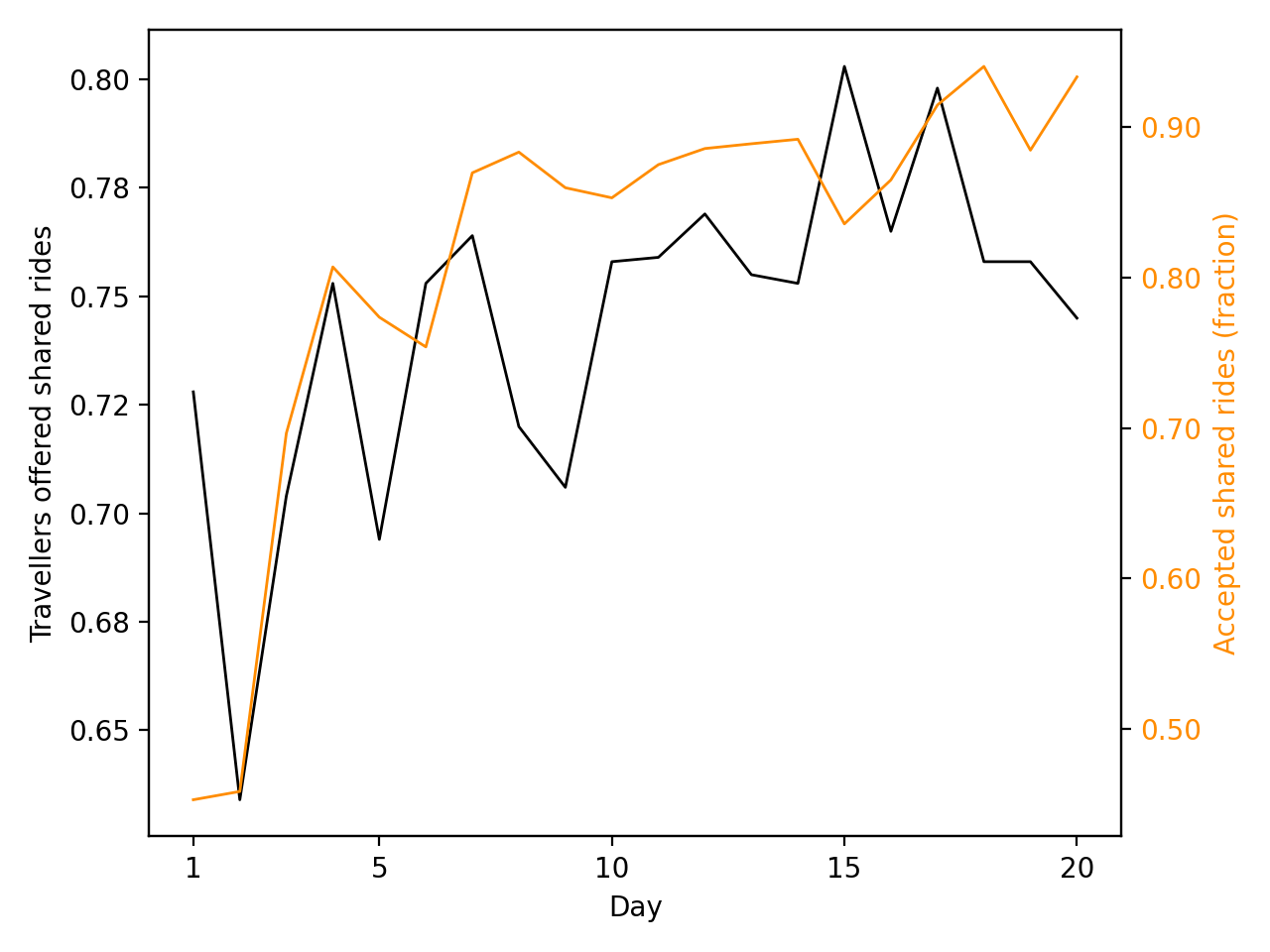}
    \caption{Fraction of travellers offered shared rides (black) and sampled acceptance rate (orange) - the fraction of realised shared rides. }
    \label{fig:sharing_acceptance}
\end{figure}

In Figure \ref{fig:sharing_acceptance}, we capture the sharing fraction and the acceptance rate. 
We plot the fraction of travellers who were offered shared rides and the actual (realised) acceptance rate (at a ride level). 
The fraction of travellers offered shared rides is within the range of $60-80\%$ during the whole service period.
We observe that, since the 10th day of operations, the fraction is in the higher portion of the range $(75-80\%)$.
Noise present in the fraction of travellers offered shared rides partly stems from incompatible travellers who have always $50\%$ participation probability.

The acceptance rate, as presented in Figure \ref{fig:sharing_acceptance}, is calculated at a ride level, i.e. presents the fraction of shared rides that are actually realised. 
Initially, a majority of shared rides ends in a fiasco. 
With time, the operator progressively meets clients' expectations and achieves around $90\%$ acceptance rate.

\subsection{Optimised discounts}\label{sec:res_fitted_discounts}
Low sharing discounts increase the profit margins while high discounts improve the likelihood that a shared ride will be accepted. 
With little knowledge of the individual value-of-time, there is a significant decision uncertainty. 
Here, we analyse the changes in the discount distribution during the learning process. 

\begin{figure}[!ht]
    \centering
    \begin{subfigure}{0.49\textwidth}
        \includegraphics[width=\textwidth]{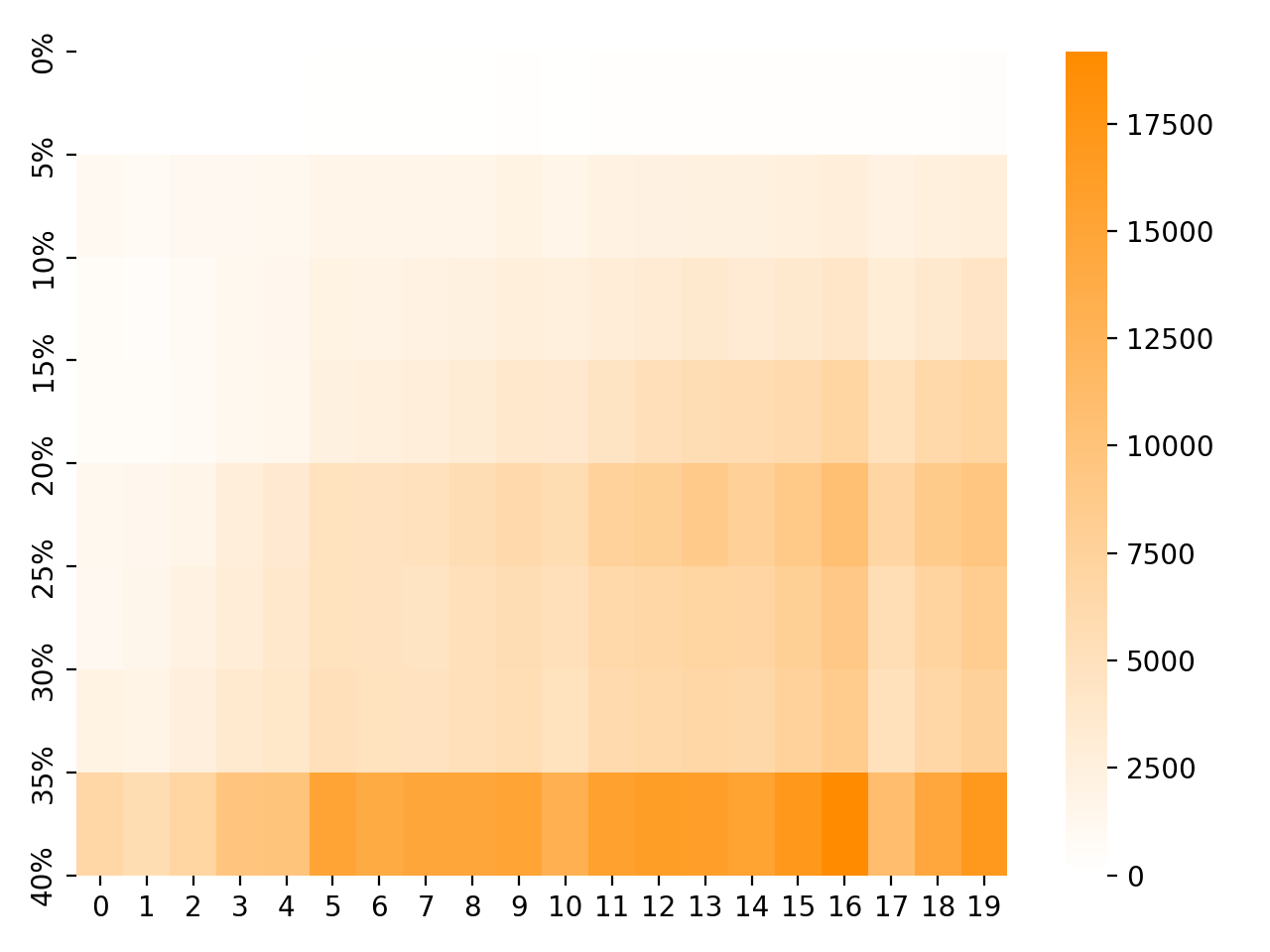}
        \caption{Shareability set}
        \label{fig:discounts_all}
    \end{subfigure}
    \begin{subfigure}{0.49\textwidth}
        \includegraphics[width=\textwidth]{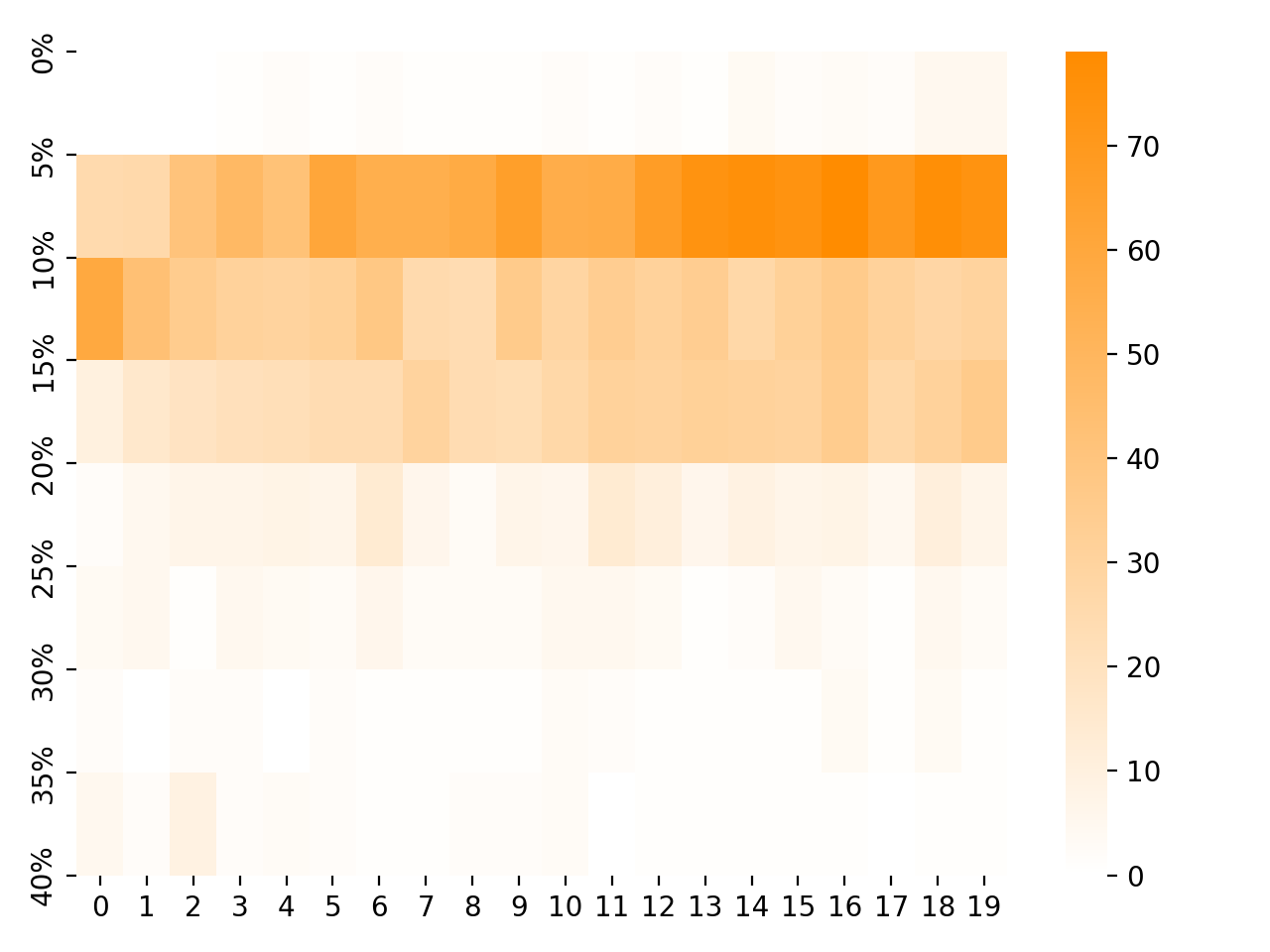}
        \caption{Offered rides}
        \label{fig:discounts_offer}
    \end{subfigure}
    \caption{Discount distribution: for all rides in the shareability set (\ref{fig:discounts_all}) and for rides selected for the offer (\ref{fig:discounts_offer}).}
    \label{fig:discounts}
\end{figure}

When we scrutinise the distribution of the sharing discounts, we can look at a two levels. 
First, all rides in the shareability set. 
Second, only rides selected for the offer in the matching process. 
In Figure \ref{fig:discounts}, we present both distributions. 

When we look at both figures, we can see significant changes in the service period. In the beginning, the operator, unsure of the individual characteristics, offers shared rides at a high discount. This protects himself from quickly discouraging travellers with higher value-of-time. As the operator learns the individual traits, he can offer shared rides with a low discount that are appreciated by travellers. Most of the offered discounts, in the latter days of the service period, are within the $5-10\%$.

We should note that most rides in the shareability set perform terribly and maximise performance under the maximum discount of $40\%$. 
It highlights the fact that groups outside the lowest value-of-time require a very significant monetary incentive to participate in most combinations. 
Also, it indicates that one can tune parameters of the shareability graph generation to less pooling-favourable which comes in favour of the real-world applications (computational complexity).

\subsection{Value-of-time estimation}\label{sec:res_class_estimation}
The crucial aspect of the efficient discount personalisation is the value-of-time class membership estimation. 
It is a backbone of our long-term optimisation. 
In this section, we look at the class estimation error.

\begin{figure}[!ht]
    \centering
    \includegraphics[width=0.8\linewidth]{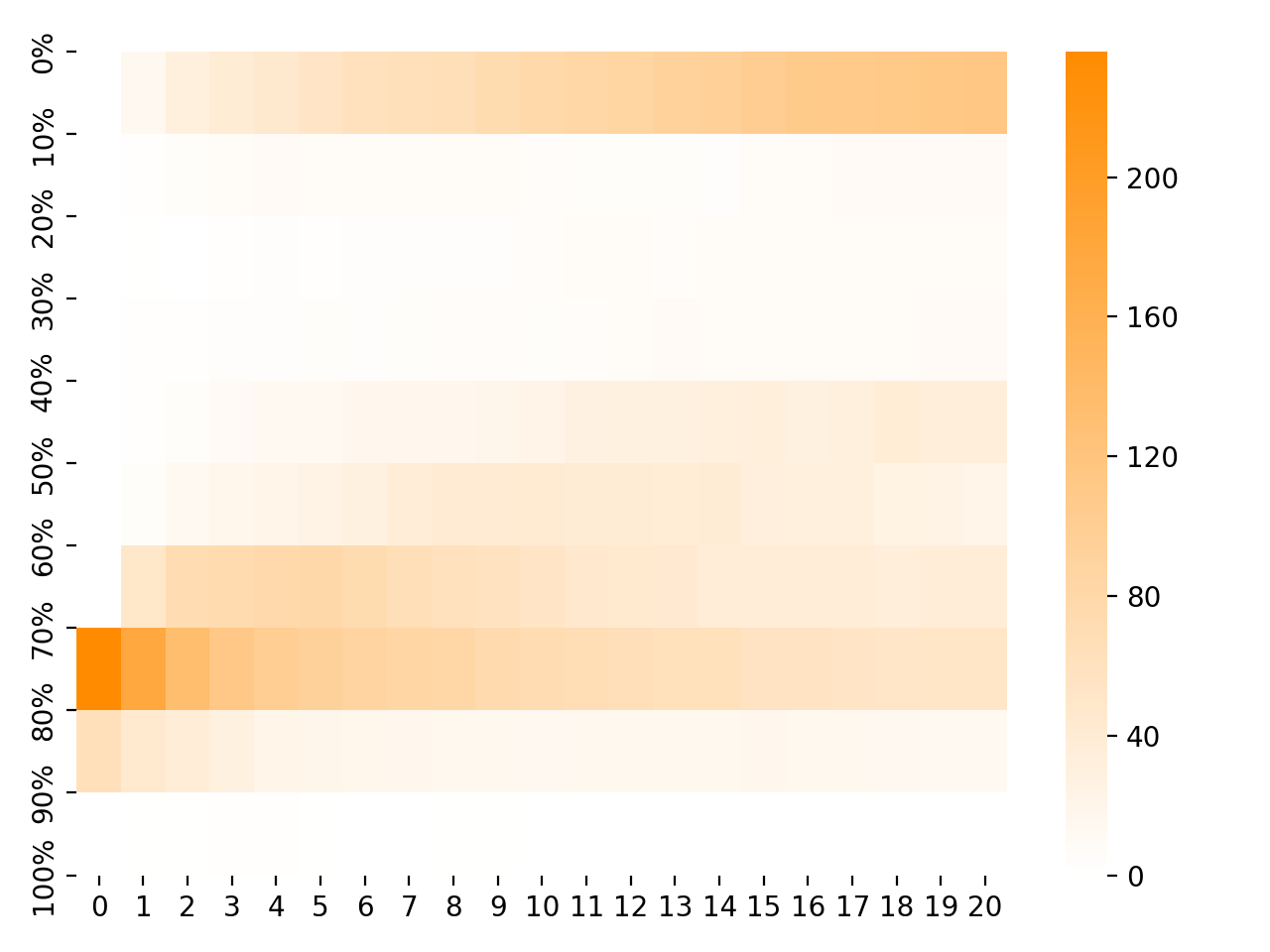}
    \caption{The class error distribution (eq. \ref{eq:vot_est_err}). The colour (right, heatmap) corresponds to the number of travellers for whom the error fit within the $10\%$ threshold level ($y$-axis) for each day ($x$-axis).}
    \label{fig:class_error}
\end{figure}

In Figure \ref{fig:class_error}, we present the class error distribution (introduced in Section \ref{sec:vot_estimation}, eq. \ref{eq:vot_est_err}) for each day. 

Initially, the class estimation error is between $70-80\%$ for travellers who are in classes of size $0.28$, $0.27$ and $0.24$ (of the total number of travellers) and $80-90\%$ for those from the smallest class of size of $0.19$. 
Travellers who experience shared rides quickly reveal their class membership. 
At the end of the service period, most travellers are estimated with a less than $10\%$ error. 
However, some travellers are never matched into shared rides and therefore the operator does not acquire any knowledge on their parameters.

\subsection{Baseline and ablation study}\label{sec:res_baseline}
To analyse whether all components of our proposed pricing policy are required to achieve a high service performance, we compare different strategies. 
We start with a flat sharing discount baseline, then introduce behavioural heterogeneity, learning, attraction and, finally, a long-term demand acquisition.

To analyse the pricing policies, we adopt the demand realised during the first day of our simulation (1st day) along with the sampled individual value-of-time. 
Additionally, we scrutinise the role of the demand attraction by including the adaptive policy with the demand acquired during the $20$ days of service (20th day).
We compute results for the cases and described in Table \ref{tab:policies}. 

\begin{table}
\centering
\caption{Pricing policies and settings.}
\label{tab:policies}
\begin{tblr}{
  width = \linewidth,
  colspec = {Q[269]Q[83]Q[244]Q[125]Q[213]},
  vline{2-6} = {-}{},
  hline{1} = {2-5}{},
  hline{2-7} = {-}{},
}
Strategy                               & Demand    & Discount policy              & Attraction Value & Individual attributes   \\
Baseline                               & 1st day   & Flat sharing discount $20\%$ & N/A              & N/A                     \\
Adaptive, 1st day                      & 1st day   & Adaptive personalised policy & 1                & Population distribution \\
Adaptive, knowledgeable                & 1st day   & Adaptive personalised policy & 1                & Learnt during $20$ days \\
Adaptive, knowledgeable, no attraction & 1st day   & Adaptive personalised policy & 0                & Learnt during $20$ days \\
Adaptive, acquired demand              & 20th day & Adaptive personalised policy & 1                & Learnt during $20$ days 
\end{tblr}
\end{table}

Our results are presented in Table \ref{tab:results}.
\begin{table}
\centering
\caption{Impact of the adaptive personalised pricing, acquired knowledge on the individual behavioural traits, attraction value on the system performance. *compared to the starting participation probability prior to operations (50\%) for the travellers joining the service}
\label{tab:results}
\begin{tblr}{
  width = \linewidth,
  colspec = {Q[219]Q[85]Q[113]Q[148]Q[210]Q[158]},
  vline{2-6} = {-}{},
  hline{2-7} = {-}{},
}
                                           & Flat pricing & Adaptive, 1st day & Adaptive, knowledgeable & Adaptive, knowledgeable, no attraction & Adaptive, acquired demand \\
Expected Profit                            & 314.55       & 363.55            & 344.24                  & 364.70                                 & \textbf{779.59}           \\
Realised Profit                            & 303.53       & 315.60            & 348.45                  & 364.33                                 & \textbf{778.01}           \\
Realised Occupancy                         & 1.25         & 1.20              & 1.39                    & 1.43                                   & \textbf{1.53}             \\
Distance Saved                             & 118.95       & 85.14             & 142.09                  & 148.28                                 & \textbf{379.78}           \\
Acceptance Rate (traveller level)          & 57\%         & 72\%              & 91\%                    & 92.5\%                                 & \textbf{96\%}             \\
Improved Participation Probability (mean)* & 4.7\%        & 4\%               & 2.2 \%                  & 1.9 \%                                 & \textbf{19.5\%}           
\end{tblr}
\end{table}


Flat sharing discount is not affected by how good travellers alignment is. 
As a result, certain travellers get an immensely attractive offer (the highest mean improvement in participation probability), yet the acceptance rate is the lowest. 
Personalised strategy improves the acceptance rate. 
Yet, without the knowledge of the individual traits, the operator opts for a conservative approach (to not discourage clients) and offers fewer shared rides. 
Once the individual characteristics are uncovered, the operator achieves a higher profit, acceptance rate, and vehicle mileage reduction. 
Removing the attraction value from the model allows for a more aggressive approach, yielding a higher instant profit, yet risking discouraging travellers.
The attraction component favours travellers crucial to the system's performance and ensures that they are compelled by an attractive offer.
Finally, the long-term adoption of our policy leads to a higher demand and the best results from the operator's, system and travellers' perspective.

\section{Conclusions}
In our study, we introduce a new pricing policy for a ride-pooling service, which is adaptive and personalised.
We show that an operator who adopts the proposed pricing policy achieves outstanding results in the long-term assessment of the service.

The policy comprises two levels: within day optimisation and day-to-day adaptation. 
In the within day optimisation, we leverage current knowledge of the individual trait (value-of-time) to fit fares which maximise total expected profit (including current and attraction value). 
In the day-to-day adaptations, based on the sharing choices (accept/reject decisions), we infer the individual behavioural classes of the travellers and predicts their satisfaction.  

In our within day optimisation, we seek personalised fares that balance two sides: travellers who accept only attractive rides and the operator who aims to maximise his profit.  
A well-fitted offer has a high acceptance rate and satisfied travellers become more likely to join the service in the following days.
Initially, the operator struggles to fit proper discounts, as he holds little knowledge of the individual preferences.
When we cannot distinguish travellers' preferences, low discounts result in a low acceptance rate, while high discounts reduce the profit margins.
To resolve this issue, we introduce the day-to-day adaptations. 
The operator scrutinises sharing decisions and infers the individual value-of-time. 
After several iterations, individual preferences are revealed, and the operator has precise estimates of the individual choice process. 
Our method leads to improved both travellers' experience and the service performance, therefore satisfying both sides of the ride-pooling service.

Our simulation shows that, to unravel the full potential of the personalised pricing in the heterogeneous population, we should apply the day-to-day adaptations. 
As we show in Section \ref{sec:res_profit}, the expected profit increases within time. 
Also, it varies less, allowing us for a more precise and reliable service performance estimations. 
In Section \ref{sec:res_satisfaction}, we show that travellers have a positive experience with shared rides, leading to a higher participation probability. 
Thanks to the adaptations, the operator offers more shared rides with a higher acceptance rate (Section \ref{sec:res_sharing_performance}). 
With increasingly precise estimates, shared rides focus around travellers who are satisfied with lower discounts (Section \ref{sec:res_fitted_discounts}). 
In Section \ref{sec:res_class_estimation}, we prove that the Bayesian inference is an effective tool for estimating individual value-of-time.
Finally, we show the role that specific model components yield to the final policy attributes (Section \ref{sec:res_baseline}).

Our adaptive pricing policy offers an interesting tool to enhance the ride-pooling offer. 
We show that the heterogeneous population requires an adaptive approach to yield the optimum results. 
To further increase applicability of the proposed method, one could consider the two-sided aspect of the market.
As we can see in the real-world adaptations, the supply-demand relations could prove instrumental in further tailoring the fares.

\section*{Acknowledgements}
This research was co-funded by the European Union's Horizon Europe Innovation Action under grant agreement No. 101103646.

This research is funded by National Science Centre in Poland program OPUS 19 (Grant Number 2020/37/B/HS4/01847).

\bibliography{references}

\end{document}